# Binary Protector: Intrusion Detection in Multitier Web Applications

C. Venkatesh[1]　　　　D. Nagaraju[2]　　　　T. Sunil Kumar Reddy[3]

[1]P.G Scholar, CSE Dept, Sir Vishveshwariah Institute of Science and Technology
[2]Assistant professor, CSE Dept, Sir Vishveshwariah Institute of Science and Technology
[3]Associate professor, CSE Dept, Sir Vishveshwariah Institute of Science and Technology

**Abstract: -** The services of internet place a key role in the daily life by enabling the in sequence from anywhere. To provide somewhere to stay the communication and management in applications the web services has stimulated to multitier design. In this multitier the web servers contain front end logic and data with database servers. In this paper, we present binary protector intrusion detection systems which designs the network behavior of user sessions across both the front-end web server and the back-end database. By examining both web and subsequent database requests, we are able to rummage out attacks that independent IDS would not be able to distinguish.

## I. INTRODUCTION

The attackers awareness has moved from attacking the front end to evade the web application vulnerabilities [1], [2], [3] in order to damage the back end database system [4] (e.g., SQL injection attacks [5], [6]). However, there is very little work being performed on multi-tier Anomaly Detection (AD) systems that generate models of network behavior for both web and database network interactions. In such multi-tier architectures, the back-end database server is often protected behind a firewall while the web servers are remotely accessible over the Internet. Unfortunately, though they are protected from direct remote attacks, the back-end systems are susceptible to attacks.

Intrusion detection systems have been widely used to protect multitier web services, such as to sense known attacks by matching changed traffic patterns or signatures [7-9]. Individually, the web IDS and the database IDS can sense uncharacteristic network traffic sent to either of them. However, these IDSs cannot detect cases wherein normal traffic is used to attack the web server and the database server. For example, if an attacker with non admin privileges can log in to a web server using normal-user access credentials, he/she can find a way to issue a privileged database query by exploiting vulnerabilities in the web server. Neither the web IDS nor the database IDS would detect this type of attack since the web IDS would merely see typical user login traffic and the database IDS would see only the normal traffic of a privileged user. This type of attack can be readily detected if the database IDS can identify that a privileged request from the web server is not associated with user-privileged access. Unfortunately, within the current multithreaded web server architecture, it is not feasible to detect or profile such causal mapping between web server traffic and DB server traffic since traffic cannot be clearly attributed to user sessions.

In this approach, it presents container based approach as shown in Fig 1.1 which is used to detect attacks in multi-tier web services. This approach can create normality models of isolated user sessions that include both the web front-end (HTTP) and back-end (File or SQL) network transactions. There is use of the container ID to accurately associate the web request with the subsequent DB queries. Thus, this guarding can build a causal mapping profile by taking both the web server and DB traffic into account.

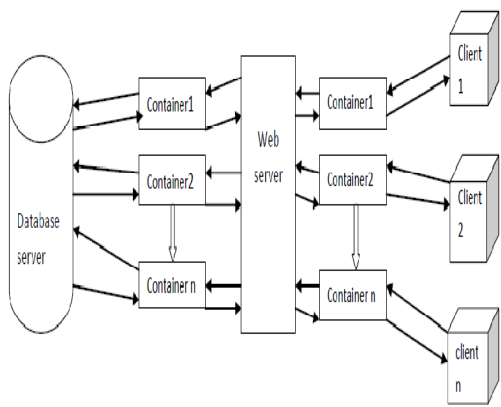

Fig 1.1: Architecture of Container

In addition to this static website case, there are web services that permit persistent back-end data modifications. These services, which we call dynamic, allow HTTP requests to include parameters that are variable and depend on user input. Therefore, the ability to model the causal relationship between the front end and back end is not always deterministic and depends primarily upon the application logic. For instance, the backend queries can vary based on the value of the parameters passed in the HTTP requests and the previous application state. Therefore, the resulting mapping between web and database requests can range from one to many, depending on the value of the parameters passed in the web request.

## 2. ANTICIPATED WORK

The breakdown structure mainly focuses on following areas –

Module 1: liable for user control; restricts illegal users.

Module 2: establishes and monitors user session.

Module 3: Checks and filters users query.

Module 4: Maps HTTP queries with equivalent SQL queries.

Module 5: Generates a log showing log of attacks.

### 2.1 Module 1: User Control

**Input:** user name and password.

**Output:** Successful or unsuccessful login.

**Algorithm:**

1. The registration form will be filled by the user.
2. Get user name and password.
3. Logs into the system.
4. Starts his new session.

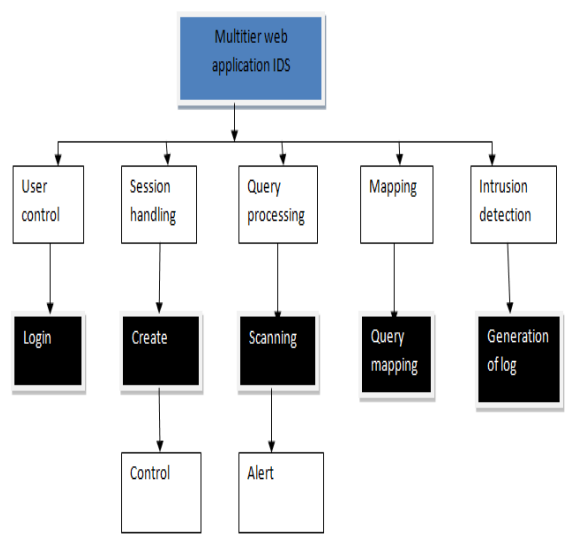

Fig 2.1 The structure of Work break Down

5. After completion of session user logs out.

The above algorithm shows how exactly the login module will provide security to the entire system to prevent unauthorized access of system. If any new user is there, wants to enter into the system then he has to fill a new user registration form. In that registration form user has to fill his personal information along with his username and password. When user clicks on save button all his information get inserted into the database.

Now this user has its own username and password. By clicking "click here to login" link he will redirect to login page. Here user will login into the system with his personal username and password. If user enters correct username and password as filled in the registration form; a "Login Successful" message will displays else if he enters wrong username or password then the system will displays "Invalid username or password message". Thus this module gives security and provides user control to the system.

**2.2 Module 2: Session Handling**

**Input:** HTTP query r and SQL query q.

**Output:** Session id for r and q in the sets ARr and AQq respectively.

**Algorithm:**

1. For each session separated traffic Ti do

2. Get different HTTP requests 'r' and DB queries 'q' in this session for each different r do

3. If r is a request to static file then

4. Add r into set EQS (Empty Query Set)

5. Else

6. If r is not in set REQ then

7. Add r into REQ

8. Append session ID i to the set ARr with r as the key

9. For each different q do

10. If q is not in set SQL then

11. Add q into SQL

12. Append session ID i to the set AQq with q as the key

Session handling module is responsible for assigning correct and unique ID to the HTTP request and equivalent SQL request. If input HTTP query is for any static data/file; means if the requested content is available at web server itself then r is added into Empty Query Set. This type of query doesn't get any kind of ID. If r is not in the set of REQ means the input query is new of arrives first time intothe system then r is added into REQ i.e. request query set. By taking r as a key session ID i is appended to the set of ARr.

Similarly for each SQL query if q is not into the set of SQL query then it is added into the SQL set. Same as above by taking q as key session ID i is appended to the set of AQq.

**2.3 Module 3: Query Processing**

**Input:** HTTP query r and SQL query q.

**Output:** Insertion of queries into different Query Sets.

**Algorithm:**

1. For each session separated traffic Ti do

2. Get different HTTP requests 'r' and DB queries 'q' in this session for each different r do

3. If r is a request to static file then

4. Add r into set EQS (Empty Query Set)

5. Else

6. If r is not in set REQ then

7. Add r into REQ

8. For each different q do

9. If q is not in set SQL then

10. Add q into SQL

   Query Processing is the module for assigning adding different requests into proper sets of query. If input HTTP query is for any static data/file; means if the requested content is available at web server itself then r is added into EQS (Empty Query Set). If r is not in the set of REQ means the input query is new of arrives first time into the system then r is added into REQ i.e. request query set. Similarly for each SQL query if q is not into the set of SQL query then it is added into the SQL set.

### 2.4 Module 4: Query Mapping

**Input:** Set of ARr, Set of AQq and Cardinality t.

**Output:** HTTP query gets mapped with equivalent SQL query.

**Algorithm:**

1. For each distinct HTTP request r in REQ do

2. For each distinct DB query q in SQL do

3. Compare the set ARr with the set AQq

4. If ARr =AQq and Cardinality(ARr) > t then

5. Found a Deterministic mapping from r to q

6. Add q into mapping model set MSr of r

7. Mark q in set SQL

8. Else

9. Need more training sessions

10. Return False

11. For each DB query q in SQL do

12. If q is not marked then

13. Add q into set NMR (No Matched Request)

14. For each HTTP request r in REQ do

15. If r has no deterministic mapping model then

16. Add r into set EQS (Empty Query Set)

17. Return True

   The user request comes to the web server in the form of HTTP request and an equivalent SQL query is generated by web server. Query mapping module maps the HTTP query with the equivalent SQL query. As we have seen the working of session handling module and query processing module. Mapping module use the output generated by these modules. A HTTP query with its ID stored in ARr set and a SQL query with its ID stored in AQq set; both are matched with each other if both ID are equal and Cardinality of ARr is greater than 1 then there is a deterministic map is found. q is then added into the matched set query and it is also marked in the set of SQL queries. After performing all training data sets if any query from the set q is not marked then that q is moved to the NMR (No Matched Request) set. Similarly for every HTTP request r; if r has no deterministic mapping then that r is added into the EQS (Empty Query Set).

### 2.5 Module 5: Intrusion Detection

**Input:** SQL query q and HTTP query r.

**Output:** Log database showing malicious query/attacks.

**Algorithm:**

1. The rule which is made for request of deterministic mapping r -> Q (Q ≠Φ), the test is taken as whether the q is a subset of a query set of the session. If so it is a valid request, and it can a marked in queries as Q. otherwise a violation is detected and treated as abnormal, and the sessions will be marked as suspicious.

2. If the rule is Empty Query Set r -> Φ, then the request is not treated as abnormal and we do not mark any database queries. Finally no intrusion will be reported.

3. For the remaining unmarked database queries, we check to see if they are in the set NMR. If so, we mark the query as such.

4. Any untested web request or unmarked database query is considered to be abnormal. If either exists within a session, then that session will be marked as suspicious.

The intrusion detection module (IDM) checks every q and r with the mapping model and then chooses that whether it is from an attacker or general user. If there is mapping found between r and q then it is a treated as valid session, otherwise it have to checks other query sets. If query r is found in Empty Query Set then it not considered as abnormal and no intrusion will be reported. For remaining unmark queries we check to see if they are in the set NMR. If so, we mark the query as such. Any query that comes directly to the database without any mapping then that session is considered as abnormal.

### 3. FUTURE SCOPE

It is promising to make some future modifications into the system; which can be make better existing system. The Intrusion detection systems can be installing on wide range of machines having different platforms and operating system. The query allowance mechanism can be made simpler by applying natural language processing (NLP); so as to convert simple English sentences into SQL queries.

Since this system works on the signature basis; each activity of intrusions is to be compared by the system earlier. New attacks are often unrecognizable by popular IDS. So there is uninterrupted race going in between detection systems and new attacks have been a challenge. at the present time Intrusion detection systems also work on the wireless networks. The latest wireless devices that break the traditional OSI layer model come with its own set of protocols for communication. So IDS must learn new communication patterns of the latest wireless technology.

### 4. CONCLUSION

An intrusion detection system which builds normality model for multitier web applications. Unlike previous techniques this technique forms container-based IDS with multiple input streams to produce alerts. There will be a technique which uses lightweight virtualization to assign session ID to a committed container which is nothing but isolated virtual computing environment. Furthermore, there will precise detection of attacks such as Hijack Future Session Attack, Privilege Escalation Attack, DB Attack and SQL Injection Attack. Log at IDS will show the details of these attacks. Also the requests which breach the normality model that will be treat as an intruder. This approach will be attempted to dynamic and static web requests with the back end file system and database queries.